# Semantic Linkage of Control Systems

**Rolf Andreas Rasenack, Karsten Wolke,
Kostyantyn Yermashov, Prof. Dr. Karl Hayo Siemsen**

FH
Oldenburg/Ostfriesland/Wilhelmshaven
Fachbereich Technik, INK
Constantiaplatz 4
26723 Emden

De Montfort University,
Software Technology Research
Laboratory
The Gateway
Leicester LE1 9BH, UK

**ABSTRACT.** Control systems are sets of interconnected hardware and software components which regulate the behaviour of processes. The software of modern control systems rises for some years by requirements regarding the flexibility and functionality. Thus the force of innovation grows on enterprises, since ever newer products in ever shorter time intervals must be made available. Associated hereby is the crucial shortening of the product life cycle, whose effects show up in reduced care of the software and the spares inventory. The aim, the concept presented here and developed in a modelling environment, is proved and ensures a minimum functionality of software components. Replacing software components of a control system verified for functionality by a framework at run-time and if necessary the software conditions will become adapted. Quintessential point of this implementation is the usage of an abstract syntax tree. Within its hierarchical structure meta information is attached to nodes and processed by the framework. With the development of the concept for semantic proving of software components the lifetime of software-based products is increased.

## 1. Introduction and Motivation

An adequate technology to develop applications is the use of modelling software components. Instead of textual keywords symbols are used for the development of applications. Within the modelling environment Neurath [Y+05] software components are ready-made. The software components are represented by icons. Behind these icons certain logic is programmed. The programmed logic is described and organized in a hierarchical structure. It is stored in a special tree, which is called abstract syntax language tree (ASLT) [W+04]. Each element or groups of elements are described by meta information. Meta information covers extended properties; they are attached as nodes within the ASLT. Meta information can be processed by software tools [Wol05]. Meta information within the ASLT permits the analysis and manipulation with programmed logic. In this article the aspect semantic linkage of software components in the specific domain of control systems is considered. With semantic linkage the functionality of software components and the relations between them is meant. During the design time (visual modelling) the requirements of software components of control systems are defined and the ready-to-use software application is generated by the model. These requirements are transferred within a special area of the software application. The result of the transfer makes linkage control of software components possible at run time.

Due to the use of new technologies, error correction and implementation of newer functionalities for software components, necessarily one or more software components of an application has to be exchanged. The new application is submitted to linkage check after the exchange of the software components at runtime. This linkage check proves the coupling of related software components and their functionality. This will be done by a framework. With the assistance of this framework the goal is pursued of proving compatibility and on a long-term basis to provide the functionality of software components. The aim is to be carried out a contribution to program software products reliable in service. The Neurath development environment used with the here presented concept fulfils advanced requirements of compatibility and automation with the assistance of the ASLT [W+04] for linkage of control systems.

## 2. Proposal

The use of modelling software components has its advantage in reducing of complexity during the software development process. The developer is able to program the application graphically (Modelling Language Neurath) [Y+05]. In addition the documentation is provided automatically to the software. Neurath Modelling Components (NMC) [Yer05] are reusable and visually manipulable software components. With their assistance a framework for semantic linkage of control systems can be developed. These Software components have defined interfaces and are combinable with other components.

This article is concerned to the question how semantic linkage between software components can be developed in a modelling environment. It is divided into two separate topics. At first the software components are built up in modelling environment Neurath. It shows the design time of the development process. Afterward the framework is discussed which gives the possibility for managing linked software components.

### 2.1. Modelled Software Components

The framework for semantic linkage of control systems will be developed by modelled software components. For this purpose the Neurath development environment will be used. The abstract syntax language tree (ASLT) and meta information have a further basic role for implementing the framework. Applications are built up from several software components. Its internal structure is shown in figure 1.

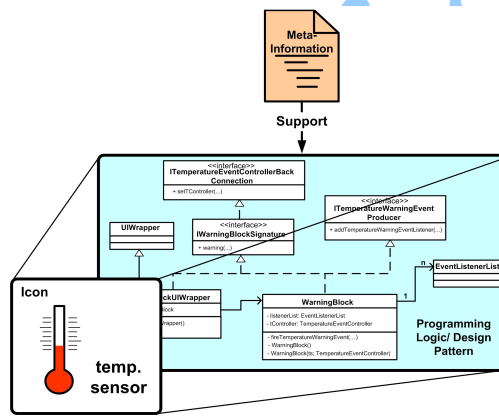

***Fig. 1:*** *A Neurath Modelling Component (NMC)*

The reason of modelling is to reduce the complexity of programming. One methodology to avoid extensive models is to articulate the semantic of applications through picot-graphic characters (isotypes). The Neurath Modelling Language (Neurath ML) [Yer06] defines a set of modelling entities represented by icons which are based on the isotype concept. The dependencies between icons are called relations and are elements of modelling entities.

With the assistance of the Neurath ML, ready-to-use components within the Neurath Environment will be developed. These software components are called Neurath Modelling Components (NMC) [Yer05]. Inside the NMC (see Figure 1) some classes and interfaces with its internal relations (e.g. inheritance) are implemented. These software components represent logic behind a Neurath icon supported by meta information.

The abstract syntax language tree is the representation of object-oriented structures (packages, classes, variables and methods), which become visible as hierarchical elements (nodes). The ASLT is the basis for variants of implementation and/or views (UML class diagram, Nassi Sheidermann diagram etc.), which are made available to the developer. Each view offers to the developer a special sight of a project. Thereby only certain parts of a project will be represented, the remaining other parts becoming invisible by folding. The ASLT is the model for the administration of hierarchic elements and particularly for the representation and/or finding of meta information, which is intended for semantic linkage of control systems [W+04, Y+04]. Meta information can be attached to the ASLT to each node. Those meta information refers then to the superior attached node. Meta information supports for example the logic of Neurath Modelling Component in the kind that the semantics of the used design pattern [J+94] as well as groups and sub-groups of associated design patterns are described. This information is read by the ASLT and becomes visible [Wol05].

A new functionality is reached by exchange of one or several software com-ponents or a discovered error is eliminated. Figure 2 shows the interaction between two NMCs which represents a complete application. By using meta information for each NMC an "interface" is provided which are represented by bubbles with the label "Linkage Control". Inside this "interface" properties have to be defined for checking the functionality of coupled software com-ponents by using the Linkage Manager. The simplest properties can be parameters of interfaces, classes or methods e. g. of an object oriented language. Software components which have to be checked about their coupling in an application are proved by this framework. After proving, the results are stored for example in a XML-file. The advantage of this procedure is to have a catalogue of proved NMCs. This is used as a cross-reference so that the information is present which NMC work together with other NMC and in which circumstances. Software components of other specific domains are recognized but not considered. In this case the XML-file works as an interface for proved NMCs. In addition it is imaginable that a database or meta information can take this part too.

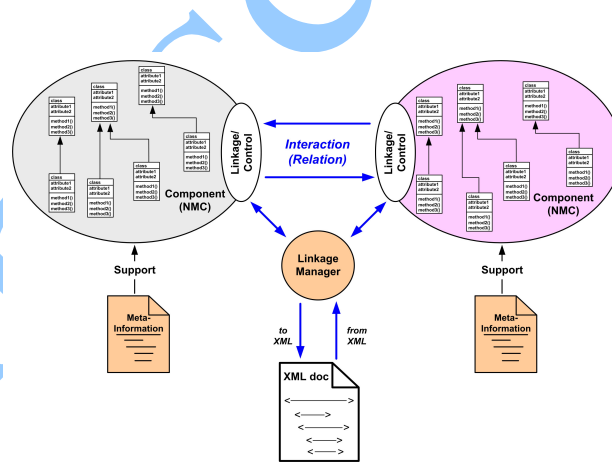

***Fig. 2:*** *Interaction between NMCs*

## 2.2 Framework for Managing Linked Software Components

The Neurath model is used for generating the application. The requirements of linked software components are implemented in the Linkage Manager. The Linkage Manager is a control mechanism of the application. It verifies the compatibility of software components and makes a roll back function if necessary. The verification of software components is activated, if at least two software components are arranged to a functional application.

The framework for semantic linkage of control systems is diagrammed in Figure 3 in principle. A software component is exchanged (software component 1, initial point) and must be merged into the existing application functionally. The exchanged software component needs a suitable software component (target point), which provides the same functionality, so that the two software components cooperate together. The Linkage Manager gets a request from the initial point to check whether a software component is present which fulfils the requirements. After that the Linkage Manager analyses (2) the corresponding software components (software component 2-5) whether appropriate functionality is present. The result of this analysis is communicated to the initial point (3). Thus as target point the software component 4 was identified.

The Linkage Manager is to be seen as a logical unit and analyses meta information. The coupling of software components, which represent classes, can be implemented for example by the use of the design pattern factory [4]. An object is produced, if a software component corresponds to the requirements of the initial point. Then the reference is returned to this component (3).

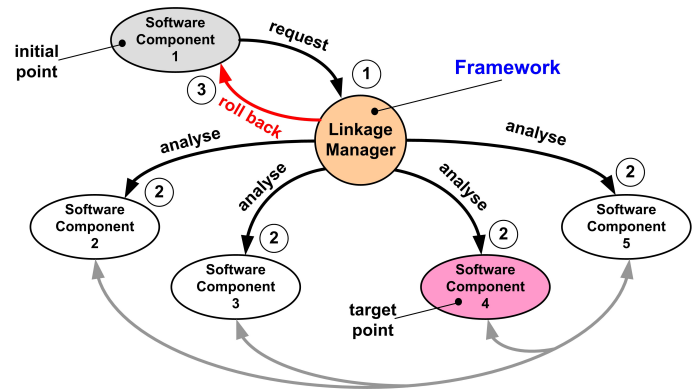

***Fig. 3:*** *Framework for Semantic Linkage of Software Components in the Specific Domain of Control*

For example with the analysis by the Linkage Manager is recognized that soft-ware components cannot work directly with each other, the so-called roll back function then is introduced. The retrieving of older revisions of software components is meant. These revisions are stored in ASLT within meta information. Due to the stored information branches or parts of ASLT can be masked. The revision and thus the functionality of the software component change dependent of going back in the revisions. It takes place an alignment up to the revision, with which the software components can co-operate again. After proofing the Linkage Mangers work is done and the application is in operation.

A condition for the successful prove of software components is their defining in a specific domain. In this case the components are specified into the specific domain for control systems. The Version Manager provides three different operation rules:
1. NMC1 works together with NMC2
2. NMC1 does not work together with NMC2
3. NMC1 works after roll back together with NMC2

The first item means that the application in which these NMC is present works without any problems. The second item gives hints that this application will not work. It includes a not successfully proofing of the roll back procedure. These software components are not compatible and another solution has to be made. During the test procedure the Linkage Manager has recognized that the software components will not work because of revision (or other) conflict. Then the roll back functionality is activated and tries to make a roll back to previous version of the software components which are working.

**Conclusion**

Meta information is used, in order to extend functionality for linkage software components. When developing applications in their components characteristics are specified, which are shown like here, used for functionality and compatibility. The software development process will be more transparent, the development time is reduced and inconsistencies are avoided by the software component based approach.

At this concept it is favourable that the proving of software components is not based on test cases and numbers, but on semantics. This is specified by the developer on component level. The developer defines interfaces, classes or methods in this context.

This concept makes a contribution to increase the lifetime of software-based products and will applied domain specific in automotive industry, automation/ home automation and embedded systems.

**References**

[J+94]   Johnson R., Gamma E., Helm R., Vlissides J. - *Design Patterns, Elements of Reusable Object Oriented Software*, Addison-Wesley, Massachusetts (USA), 1994, ISBN 0-201-63361-2

[Wol05]  Wolke K. - *Meta-Information and its Processing*, Fachhochschule Oldenburg/Ostfriesland/Wilhelmshaven, Standort Emden (D), Fachbereich Technik and STRL, DeMontfort University Leicester (GB), 2006, http://www.karsten-wolke.de/public/aslt/ASLTMetaData.pdf

[W+04]   Wolke K., Yermashov K., Siemsen K.H., Rasenack R.A., Abt C. - *Abstract Syntax Trees for Source Code Management*, Toolbox Magazin (D), September/Oktober 2004

[Yer05]  Yermashov K. - *Modeling Language Neurath Tutorial*, Fachhochschule Oldenburg/Ostfriesland/Wilhelmshaven, Standort Emden (D), Fachbereich Technik, August 2005

[Yer06]   Yermashov K. - *Framework for Domain-Specific Semantic Modelling of Software. Specification of Security Requirements in a Security Layer*, Fachhochschule Oldenburg/Ostfriesland/ Wilhelmshaven, Standort Emden (D), Fachbereich Technik and STRL, DeMontfort University Leicester (GB), 2006

[Y+05]   Yermashov K., Wolke K., Abt C., Rasenack R.A., Siemsen K.H. - *Integrated Modeling Environment with Modeling Language Neurath*, Fachhochschule Oldenburg/Ostfriesland/ Wilhelmshaven, Standort Emden (D), Fachbereich Technik, 2005

[Y+04]   Yermashov K., Wolke K., Siemsen K.H., Abt C., Rasenack R.A. - *From Diagram to Source Code*, Toolbox Magazin (D), Mai/June 2004